\documentclass[twocolumn]{jpsj3} 

\usepackage{color}

\title{
Photoexcitation-Energy-Dependent Transition Pathways from a Dimer Mott Insulator to a Metal
}

\author{
Kenji \textsc{Yonemitsu}$,^{1,2,3}$\thanks{E-mail address: kxy@ims.ac.jp} 
Satoshi \textsc{Miyashita}$,^{1,4}$\thanks{Present address: JST, Tokyo 102-0075, Japan} and 
Nobuya \textsc{Maeshima}$^{5,6}$
}

\inst{
$^{1}$Institute for Molecular Science, Okazaki, Aichi 444-8585, Japan \\
$^{2}$Department of Functional Molecular Science, Graduate University for Advanced Studies, Okazaki, Aichi 444-8585, Japan \\
$^{3}$JST, CREST, Tokyo 102-0075, Japan \\
$^{4}$Institute for Materials Research, Tohoku University, Sendai, Miyagi 980-8577, Japan \\
$^{5}$Institute of Materials Science, University of Tsukuba, Tsukuba, Ibaraki 305-8573, Japan \\
$^{6}$Center for Computational Sciences, University of Tsukuba, Tsukuba, Ibaraki 305-8577, Japan 
}

\abst{
We theoretically study pump-photon-energy-dependent pathways of a photoinduced dimer-Mott-insulator-to-metal transition, on the basis of numerical solutions to the time-dependent Schr\"odinger equation for the exact many-body wave function of a two-dimensional three-quarter-filled extended Peierls-Hubbard model. When molecular degrees of freedom inside a dimer are utilized, photoexcitation can weaken the effective interaction or increase the density of photocarriers. In the organic dimer Mott insulator, $ \kappa $-(BEDT-TTF)$_2$Cu[N(CN)$_2$]Br, the intradimer and the interdimer charge-transfer excitations have broad bands that overlap with each other. Even in this disadvantageous situation, the photoinduced conductivity change depends largely on the pump photon energy, confirming the two pathways recently observed experimentally. The characteristic of each pathway is clarified by calculating the modulation of the effective interaction and the number of carriers involved in low-energy optical excitations. The pump-photon-energy-dependent pathways are confirmed to be realized from the finding that, although the effective interaction is always and slowly weakened, the introduction of carriers is sensitive to the pump-photon energy and proceeds much faster. 
}

\kword{
photoinduced phase transition, metal-insulator transition, dimer Mott insulator, electron-phonon interaction
}

\begin{document}
\maketitle

\section{Introduction}

Nonequilibrium properties of strongly correlated electron systems attract much attention. Among them, photoinduced phase transition dynamics in different groups of materials are now deeply and extensively investigated from experimental and theoretical aspects. \cite{tokura_jpsj06} Their ultrafast and nonequilibrium characteristics are important, which we can take advantage of to explore novel functions. Most of their cooperative characteristics are explained on the basis of itinerant electron models. \cite{yonemitsu_jpsj06} 

Photoexcitation energy dependence is always a key issue. If the relationship between the density of absorbed photons and the amount of photoinduced reflectivity change (such as efficiency and nonlinearity) depends largely on the photoexcitation energy, it will give a useful hint for optical control in the future. For instance, in neutral-ionic transitions in TTF-CA (TTF=tetrathiafulvalene, CA=chloranil), this relationship depends on the photoexcitation energy \cite{suzuki_prb99,huai_jpsj00} and the direction of the transition. \cite{okamoto_prb04,yonemitsu_prb06} This information contributes to the knowledge of its transition pathway. \cite{okamoto_prb04,uemura_prl10}

If the transition pathway depends on the photoexcitation energy, it will directly lead to the optical control of electronic properties. For instance, Mott insulators are known to be generally converted into metals either by weakening the effective on-site repulsion or by introducing carriers. \cite{imada_rmp98} Photoexcitation is often utilized to introduce carriers into one-dimensional Mott insulators and induce metallic conductivity. \cite{iwai_prl03,okamoto_prl07} One-dimensional Mott insulators are, however, special in that infinitesimal on-site repulsion leads to the insulating ground state, \cite{solyom_ap79,schulz_ijmpb91} so that only the introduction of carriers causes metallic conductivity. \cite{maeshima_jpsj05}

In two-dimensional Mott insulators, the situation is thus drastically different from the above. For instance, the Mott insulating ground state in copper oxides is converted into a superconducting state by chemical doping, \cite{imada_rmp98,uchida_prb91} whereas that in organic salts is converted into a superconducting state by applying chemical pressure to weaken the effective on-site repulsion relative to the bandwidth. \cite{kanoda_hyperfine97,kanoda_jpsj06} Thus, their photoexcitation may be able to induce a Mott-insulator-to-metal transition via one of these pathways, which generally depends on the photoexcitation energy. Indeed, this has been suggested to be achieved in deuterated $ \kappa $-($ d $-BEDT-TTF)$_2$Cu[N(CN)$_2$]Br [BEDT-TTF=bis(ethylenedithio)-tetrathiafulvalene]. \cite{kawakami_prl09} 

In this paper, we discuss how the effective interaction and the carrier density are modulated by photoexcitation in the dimer Mott insulator. It is essential to take molecular degrees of freedom inside a dimer into account. The organic (BEDT-TTF)$_2$X salts basically have a three-quarter-filled band. Because of strong dimerization of BEDT-TTF molecules, the $ \kappa $-type salts are generally assumed to have the completely filled band originating from bonding orbitals and the half-filled band originating from antibonding orbitals, which are well separated in energy space. \cite{kino_jpsj96} Thus, they are often regarded as half-filled band systems \cite{kandpal_prl09,nakamura_jpsj09} with a Mott insulating ground state. 

It is well known that the effective on-site repulsion strength in such a system is given by the transfer integral between molecular orbitals inside a dimer in the limit of strong on-site repulsion on a molecular orbital. \cite{kanoda_hyperfine97} Its modulation by photoexcitation is described in a three-quarter-filled-band model, i.e., with two molecular orbitals per dimer. This transfer integral is expected to depend sensitively on the distance and the relative orientation of these molecules. This sensitivity would be responsible for the chemical-pressure-temperature phase diagram \cite{kanoda_jpsj06} in addition to that of interdimer transfer integrals. 

The effects of photoexcitation are not so simply described. The resultant charge-transfer (CT) process alters the stable molecular configuration and thus modifies the effective on-site repulsion strength. It also generally introduces electrons and holes. These two effects are always realized, and their relative weights depend on the photoexcitation energy. In $ \kappa $-(BEDT-TTF)$_2$Cu[N(CN)$_2$]Br, both the intradimer and interdimer CT excitations have broad bands, which overlap to a large extent. \cite{faltermeier_prb07,dumm_prb09} Therefore, a single effect is not readily realized simply by tuning the photoexcitation energy. 

Motivated by this fact, we employ a three-quarter-filled extended Peierls-Hubbard model on an anisotropic triangular lattice. Phonons modulating the effective on-site repulsion are treated quantum mechanically, so that the intradimer and interdimer CT excitations have overlapped bands even in small clusters. Even under this disadvantageous condition, the selection of the transition pathway is shown to be realized in numerical solutions to the time-dependent Schr\"odinger equation for the exact many-electron-phonon wave function. 

\section{Three-Quarter-Filled Model for Dimer Mott Insulator}

\begin{figure}
\includegraphics[height=6cm]{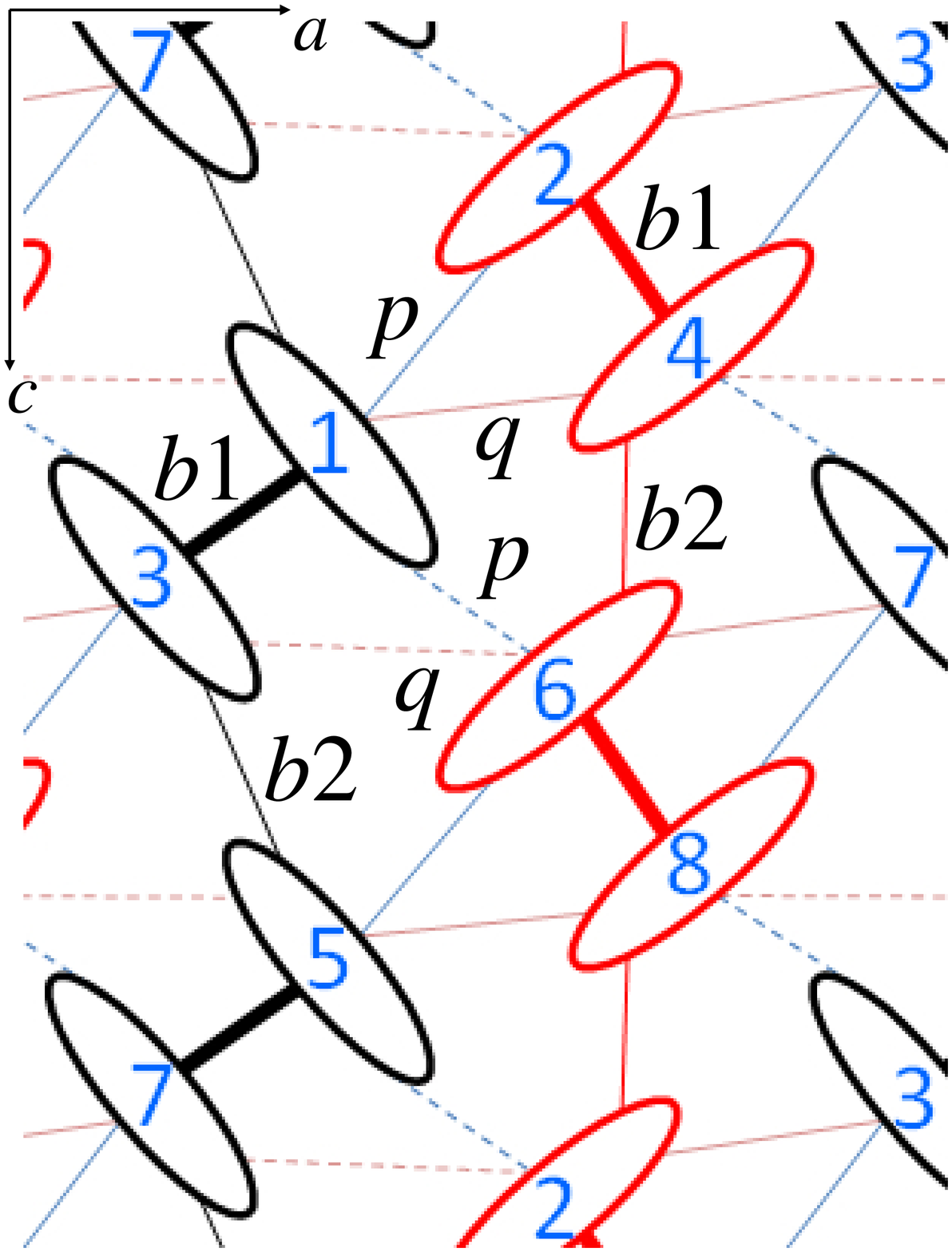}
\caption{(Color online) 
Anisotropic triangular lattice for dimer Mott insulator.
\label{fig:lattice}}
\end{figure}

The model we use is the three-quarter-filled extended Peierls-Hubbard model on the anisotropic triangular lattice shown in Fig.~\ref{fig:lattice}, 
\begin{eqnarray}
H & = & 
\sum_{\langle ij \rangle \sigma} \left\{
\left[ t^{(0)}_{ij} -g_{ij} \left( b_{ij} + b^\dagger_{ij} \right) \right] 
c^\dagger_{i\sigma} c_{j\sigma} +\mbox{H.c.}
\right\}
\nonumber \\ & & 
+U\sum_i n_{i\uparrow} n_{i\downarrow}
+\sum_{\langle ij \rangle} V_{ij} n_i n_j
+\sum_{\langle ij \rangle } \omega_{ij} b^\dagger_{ij} b_{ij}
\;, \label{eq:model}
\end{eqnarray}
where $ c^\dagger_{i\sigma} $ creates an electron with spin $ \sigma $ at site $ i $, $ n_{i\sigma} $=$ c^\dagger_{i\sigma} c_{i\sigma} $, and $ n_i $=$ \sum_\sigma n_{i\sigma} $. The parameter $ U $ represents the on-site Coulomb repulsion, and the intersite Coulomb repulsion $ V_{ij} $ is assumed to be $ V_{ij}=V_0/\! \mid \! \mbox{\boldmath $ r $}_{i}-\mbox{\boldmath $ r $}_{j} \! \mid $ for the four types of pairs $ \langle ij \rangle $ denoted by $ b1 $, $ b2 $, $ p $, and $ q $ in Fig.~\ref{fig:lattice}. Here, the intermolecular distance is taken from ref.~\citen{mori_bcsj99}. The intradimer $ V_{ij} $=$ V_{b1} $ is the largest among $ V_{b1} $, $ V_{b2} $, $ V_{p} $, and $ V_{q} $. 

The operator $ b^\dagger_{ij} $ creates a phonon, which is assumed to modulate only the intradimer transfer integral $ t_{b1} $. The parameters $ g_{ij} $ and $ \omega_{ij} $ are the corresponding electron-phonon coupling strength and the bare phonon energy, $ g_{b1} $ and $ \omega_{b1} $, respectively. 
The transfer integrals $ t_{ij} $ for $ \kappa $-($ d $-BEDT-TTF)$_2$Cu[N(CN)$_2$]Br are estimated from the extended H\"uckel calculation: \cite{watanabe_sm99} 
$ t_{b1} \simeq t^{(0)}_{b1}-2g_{b1}^2/\omega_{b1} $=$-$0.2756, 
$ t_{b2} $=$-$0.1047, $ t_{p} $=$-$0.1115, and $ t_{q} $=0.0404 in units of eV. 
Here, the transfer integrals that are not modulated by phonons are denoted without the superscript (0). 
We set $ U $=0.8 and $ V_{b1} $=0.3. 
Here, we employ the transfer integrals $ t_{ij} $ for $ \kappa $-($ d $-BEDT-TTF)$_2$Cu[N(CN)$_2$]Br because this insulating material is located in the close vicinity of the metal-insulator phase boundary. Because it is impossible to extrapolate from the results of the present finite-size system to those in the thermodynamic limit, we cannot judge whether or not the present parameter set of $ t_{ij} $, $ U $, and $ V_{b1} $ really corresponds to an insulating ground state in this limit. However, the present results and conclusions are not modified by the details in these parameters, so that we use this parameter set. 
As for phonons, we take a strong electron-phonon coupling $ g_{b1} $=0.06 and a high phonon energy $ \omega_{b1} $=0.05 to make the intradimer and interdimer CT bands overlap to a large extent. 

The time-dependent Schr\"odinger equation is solved for the many-electron-phonon wave function on the cluster of $ N $=8 sites with a periodic boundary condition (Fig.~\ref{fig:lattice}) and with the number of phonons restricted to a maximum of three at any $ b1 $ bond. It is numerically solved by expanding the exponential evolution operator with time slice $ dt $=0.02 eV$^{-1}$ to the 15th order and by checking the conservation of the norm. This is basically explained in ref.~\citen{yonemitsu_prb09} for the many-electron wave function, and it is now extended to the many-electron-phonon wave function $ \mid \! \psi(t) \rangle $. Photoexcitation is introduced through the Peierls phase, 
\begin{equation}
c^\dagger_{i\sigma} c_{j\sigma}
\rightarrow
e^{(ie/\hbar c) \mbox{\boldmath $ \delta $}_{ij} \cdot \mbox{\boldmath $ A $}(t)}
c^\dagger_{i\sigma} c_{j\sigma}
\;, \label{eq:Peierls}
\end{equation}
with $ \mbox{\boldmath $ \delta $}_{ij}=\mbox{\boldmath $ r $}_{j}-\mbox{\boldmath $ r $}_{i} $. The time-dependent vector potential $ \mbox{\boldmath $ A $}(t) $ for a pulse of an oscillating electric field is given by 
\begin{equation}
\mbox{\boldmath $ A $}(t)=
\frac{\mbox{\boldmath $ F $}}{\omega_\mathrm{pmp}}
\cos(\omega_\mathrm{pmp} t)
\frac1{\sqrt{2\pi}T_\mathrm{pmp}}
\exp \left( -\frac{t^2}{2 T_\mathrm{pmp}^2} \right)
\;, \label{eq:vector_potential}
\end{equation}
where the electric field amplitude $ \mbox{\boldmath $ F $}$ is set parallel to the $ c $-axis (the $ c $-component is denoted by $ F_c $), the pulse width is $ T_\mathrm{pmp} $=35 eV$^{-1}$=23 fs, and $ \omega_\mathrm{pmp} $ is the excitation energy. 

The transient optical conductivity $ \sigma'(\omega_\mathrm{prb},t) $ is calculated as before, \cite{matsueda_jpsj07,onda_prl08} 
\begin{equation}
\sigma'(\omega_\mathrm{prb},t) = -\frac{1}{N\omega_\mathrm{prb}} \mathrm{Im} 
\langle \psi(t) \mid \mbox{\boldmath $ j $}
\frac{1}{\omega_\mathrm{prb} + i \epsilon + E - H}
\mbox{\boldmath $ j $} \mid \psi(t) \rangle
\;, \label{eq:conductivity}
\end{equation}
where $ \mbox{\boldmath $ j $} \equiv - \partial H/\partial \mbox{\boldmath $ A $} $ is the current operator, $ \epsilon $ is a peak-broadening parameter set at 0.005, and $ E = \langle \psi(t) \mid H \mid \psi(t) \rangle $. 

\section{Intra- and Interdimer CT Excitations \label{sec:intra_inter}}

Here we discuss the optical modulation of the effective on-site repulsion through intradimer and interdimer CT excitations. As a first step, we consider an isolated dimer consisting of two molecular orbitals overlapped with a transfer integral $ t_{b1} $. One-hole states consist of the bonding and antibonding states with energies $ \pm \! \mid \! t_{b1} \! \mid $. Two-hole states consist of three singlet states with energies $ U $, $ (U+V_{b1})/2 \pm \sqrt{ (U-V_{b1})^2/4 + 4 t_{b1}^2 } $ ($ \simeq U $, $ V_{b1} $ for $ \! \mid \! t_{b1} \! \mid \ll U $, $ V_{b1} $), and one triplet state with energy $ V_{b1} $. The effective on-site Coulomb energy, defined as $ E_2+E_0-2E_1 $ with $ E_n $ for the energy of the lowest $ n $-hole state, is given by 
\begin{equation}
U_\mathrm{dim} = \frac{ U+V_{b1} }{2} - \sqrt{ \left(\frac{ U-V_{b1} }{2}\right)^2 + 4 t_{b1}^2 } +2 \! \mid \! t_{b1} \! \mid
\;. \label{eq:udim}
\end{equation}
It becomes $ U_\mathrm{dim} \simeq 2 \! \mid \! t_{b1} \! \mid \! + V_{b1} $ in the limit of $ \! \mid \! t_{b1} \! \mid \ll U $, $ V_{b1} $. 

\begin{figure}
\includegraphics[height=6cm]{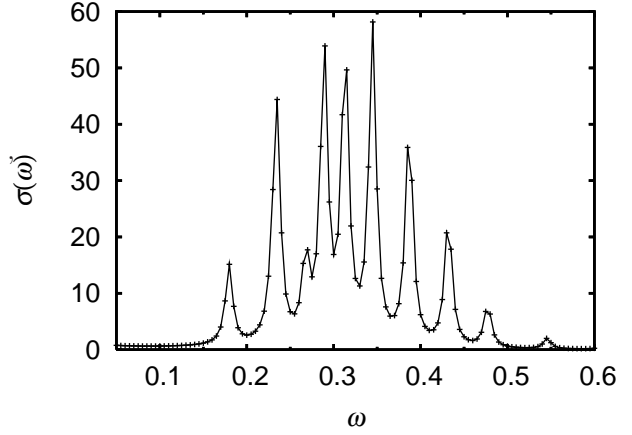}
\caption{
Optical conductivity with polarization parallel to the $ c $-axis.
\label{fig:conductivity}}
\end{figure}

In this strong-coupling limit, the intradimer CT excitation between the bonding and the antibonding states costs $ 2 \! \mid \! t_{b1} \! \mid $, while the interdimer CT excitations cost $ 2 \! \mid \! t_{b1} \! \mid \! + U $ and $ 2 \! \mid \! t_{b1} \! \mid \! + V_{b1} $. Therefore, the interdimer CT excitations require higher energies in this limit. However, the energy of the intradimer CT excitation is lowered by the second-order perturbation with respect to the interdimer transfer integrals, whereas those of the interdimer CT excitations are lowered by the first-order perturbation. Using realistic values for transfer integrals in $ \kappa $-(BEDT-TTF)$_2$Cu[N(CN)$_2$]Br, we actually find that the intradimer CT excitation has a higher energy than the interdimer CT excitations, as already assigned in the experimental works. \cite{faltermeier_prb07} The optical conductivity is calculated by substituting the ground state $ \mid \psi_0 \rangle $ for $ \mid \psi(t) \rangle $ in eq.~(\ref{eq:conductivity}), as shown in Fig.~\ref{fig:conductivity}. 
As clearly shown, the charge gap is about 0.18. Because the bare phonon energy is set to be $ \omega_{b1} $=0.05, the $ \sigma(\omega) $ spectrum has peaks at an interval of about 0.05. The peaks are not exactly located at even intervals because they correspond to phonon shake-off processes associated with different electronic excitations: intradimer and interdimer CT excitations. In other words, the intradimer and interdimer CT bands are largely overlapped by phonon excitations. In experimental situations, different modes of phonons with much smaller energies contribute to the band broadening, so that the spectrum consists of a smooth curve. In the calculations performed below, the excitation energy $ \omega_\mathrm{prb} $ is set to be an off-resonant energy. As demonstrated below, the excitations around $ \omega $=0.3 are mainly due to interdimer CT excitations, whereas the excitations on the high-energy side are mainly due to intradimer CT excitations. Taking this large overlap between them, the picture based on the dimer Mott insulator may not be so accurate. At least, the effective on-site repulsion strength would deviate from the value in the strong-coupling limit. 

The way in which the effective on-site repulsion is optically modulated is described rather well in the classical picture for phonons. The terms involving phonon operators in the Hamiltonian are then approximated to be
\begin{eqnarray}
H_\mathrm{ph} & \simeq & 
\sum_{\langle ij \rangle} \left[
 -\alpha_{b1} u_{ij} \sum_\sigma \left(
c^\dagger_{i\sigma} c_{j\sigma} + c^\dagger_{j\sigma} c_{i\sigma}
\right)
\right. \nonumber \\ & & \left.
+ \frac{K_{b1}}{2} 
\left( u_{ij}^2 + \frac{\dot{u}_{ij}^2}{\omega_{b1}^2} \right) \right]
\;, \label{eq:classical}
\end{eqnarray}
where sites $ i $ and $ j $ correspond to two molecules in a dimer, the displacement $ u_{ij} $ is proportional to the classical analog of $ ( b_{ij} + b^\dagger_{ij} ) $, and $ \alpha_{b1}^2/ K_{b1} $ corresponds to the coupling strength. The force applied to $ u_{ij} $ is thus given by 
\begin{eqnarray}
-\langle \frac{\partial H}{\partial u_{ij}} \rangle & = & 
\alpha_{b1} \sum_\sigma 
\langle 
c^\dagger_{i\sigma} c_{j\sigma} + c^\dagger_{j\sigma} c_{i\sigma}
\rangle - K_{b1} u_{ij}
\nonumber \\ & = & 
\alpha_{b1} \sum_\sigma 
\langle 
b^\dagger_{l\sigma} b_{l\sigma} - a^\dagger_{l\sigma} a_{l\sigma}
\rangle - K_{b1} u_{ij}
\;, \label{eq:force}
\end{eqnarray}
where we rewrite the electron operators as
\begin{equation}
b_{l\sigma} , a_{l\sigma} = (c_{i\sigma} \pm c_{j\sigma} )/\sqrt{2}
\label{eq:bonding}
\end{equation}
in terms of the bonding and antibonding orbitals in dimer $ l $ containing molecules $ i $ and $ j $. It can be shown, in a straightforward manner, that the kinetic term of eq.~(\ref{eq:model}) consists of $ b^\dagger_{k\sigma} b_{l\sigma} $ and $ a^\dagger_{k\sigma} a_{l\sigma} $ operators only, whereas the current operator $ \mbox{\boldmath $ j $} $ consists of $ a^\dagger_{k\sigma} b_{l\sigma} $ and $ b^\dagger_{k\sigma} a_{l\sigma} $ operators only. Without Coulomb interactions and without photoexcitation, the number of electrons in the bonding orbitals and that in the antibonding orbitals were conserved. 
Because the number of electrons in the bonding orbitals is larger than that in the antibonding orbitals, any photoexcitation reduces 
$ \sum_\sigma \langle 
b^\dagger_{l\sigma} b_{l\sigma} - a^\dagger_{l\sigma} a_{l\sigma}
\rangle $, 
which causes force to be applied to $ u_{ij} $ in the negative direction, reduces the magnitude of the intradimer transfer, and consequently weakens the effective on-site repulsion, $ U_\mathrm{dim} $. This situation is numerically confirmed below.

\section{Modulation of Effective Interaction and Spectral Weight after Photoexcitation \label{sec:modulation}}

In general, it is difficult to compare numerical results with the experimental observation when model calculations are based on exact many-electron-phonon wave functions, because of the limitation of small sizes of clusters and small numbers of allowed phonon excitations. The present cluster does not show a metal-insulator transition because a finite charge gap always exists. However, we can evaluate the weakening of the effective interaction and the carrier-induced spectral-weight transfer, both of which lead to an insulator-to-metal transition in the thermodynamic limit. 

\begin{figure}
\includegraphics[height=6cm]{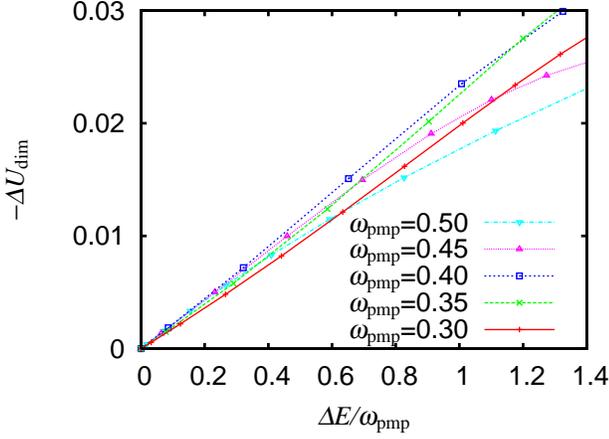}
\caption{(Color online) 
Modulation of effective on-site repulsion $ -\Delta U_\mathrm{dim} $, as a function of the number of absorbed photons $ \Delta E / \omega_\mathrm{pmp} $, for $ \omega_\mathrm{pmp} $=0.30, 0.35, 0.40, 0.45, and 0.50.
\label{fig:Udim}}
\end{figure}

In order to see the modulation of the effective on-site repulsion $ U_\mathrm{dim} $ in a direct manner, we calculate the expectation value of the displacement $ \langle b_{ij} + b^\dagger_{ij} \rangle $ up to $ t $=450. Its maximum decrement, $ -\Delta \langle b_{ij} + b^\dagger_{ij} \rangle $, gives the maximum decrement in  $ U_\mathrm{dim} $, $ -\Delta U_\mathrm{dim} $. We evaluate it, using eq.~(\ref{eq:udim}) and the equation $ t_{b1} = t^{(0)}_{b1} -g_{b1} \left( b_{ij} + b^\dagger_{ij} \right) $ with $ t^{(0)}_{b1} < 0 $. We vary the electric field amplitude $ F_c $ and calculate the increment in the total energy $ \Delta E $ divided by $ \omega_\mathrm{pmp} $, which corresponds to the number of absorbed photons. We show $ -\Delta U_\mathrm{dim} $ in Fig.~\ref{fig:Udim} as a function of $ \Delta E / \omega_\mathrm{pmp} $ for different $ \omega_\mathrm{pmp} $. 
As discussed in \S~\ref{sec:intra_inter}, the current operator $ \mbox{\boldmath $ j $} $ consists of $ a^\dagger_{k\sigma} b_{l\sigma} $ and $ b^\dagger_{k\sigma} a_{l\sigma} $ operators ($ k = l $ for intradimer and $ k \neq l $ for interdimer CT processes). Both the intradimer and interdimer CT processes basically reduce $ \sum_{l\sigma} \langle 
b^\dagger_{l\sigma} b_{l\sigma} - a^\dagger_{l\sigma} a_{l\sigma}
\rangle $ by two. The force applied to phonons is therefore similar between these processes. As a consequence, the ratio of $ -\Delta U_\mathrm{dim} $ to $ \Delta E / \omega_\mathrm{pmp} $ is similar in the range of $ 0.3 < \omega_\mathrm{pmp} < 0.5 $. The effective on-site repulsion is confirmed to be weakened to a similar extent irrespective of whether charge is transferred mainly within a dimer or mainly between dimers. 

The number of carriers involved in the optical excitations up to $ \omega_\mathrm{prb} $ is known to be proportional to the spectral weight obtained by the integration of the conductivity over $ 0 < \omega < \omega_\mathrm{prb} $ below the charge gap. This quantity has been measured, both in equilibrium \cite{uchida_prb91} and after photoexcitation. \cite{iwai_prl03} Here, we calculate the increment in the conductivity $ \Delta \sigma'(\omega_\mathrm{prb},t) = \sigma'(\omega_\mathrm{prb},t) - \sigma'(\omega_\mathrm{prb},-150) $, time-average it over $ 150 < t < 750 $ $ \Delta \sigma'(\omega_\mathrm{prb}) = (1/600)\int_{150}^{750} \Delta \sigma'(\omega_\mathrm{prb},t) dt $, and integrate it over $ 0 < \omega < \omega_\mathrm{prb} $, 
\begin{equation}
N(\omega_\mathrm{prb}) = \int_0^{\omega_\mathrm{prb}} \Delta \sigma'(\omega) d\omega
\;.
\end{equation}

\begin{figure}
\includegraphics[height=12cm]{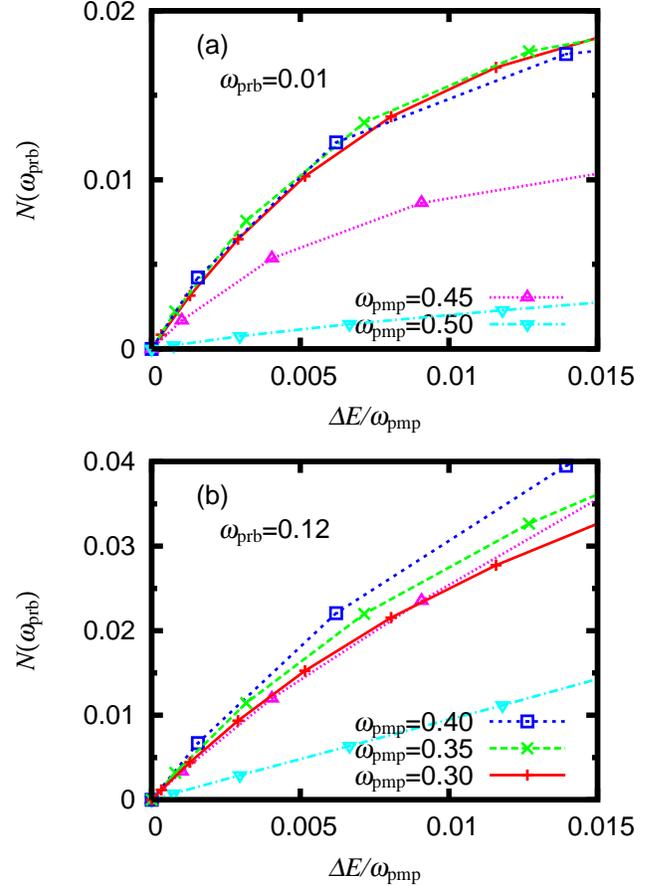}
\caption{(Color online) 
Increment in conductivity time-averaged and integrated over $ 0 < \omega < \omega_\mathrm{prb} $, $ N(\omega_\mathrm{prb}) $, for (a) $ \omega_\mathrm{prb} $=0.01 and (b) $ \omega_\mathrm{prb} $=0.12, as a function of the number of absorbed photons $ \Delta E / \omega_\mathrm{pmp} $, for $ \omega_\mathrm{pmp} $=0.30, 0.35, 0.40, 0.45, and 0.50.
\label{fig:int_sgm}}
\end{figure}

We show $ N(\omega_\mathrm{prb}) $ in Fig.~\ref{fig:int_sgm} as a function of $ \Delta E / \omega_\mathrm{pmp} $ for different $ \omega_\mathrm{pmp} $. 
The energy $ \omega_\mathrm{prb} $ is set at 0.01 in Fig.~\ref{fig:int_sgm}(a) and at 0.12 in Fig.~\ref{fig:int_sgm}(b) below the charge gap of 0.18 in the ground state. To maintain the numerical accuracy, we use weaker electric field amplitudes $ F_c $ so that $ \Delta E / \omega_\mathrm{pmp} $ are smaller than those used for the evaluation of $ -\Delta U_\mathrm{dim} $. Although the quantity $ N(\omega_\mathrm{prb}) $ increases with $ \Delta E / \omega_\mathrm{pmp} $ for any $ \omega_\mathrm{pmp} $, its rate depends largely on $ \omega_\mathrm{pmp} $. For any $ \omega_\mathrm{prb} $ below the charge gap, $ N(\omega_\mathrm{prb}) $ increases rapidly for $ \omega_\mathrm{pmp} $=0.3, 0.35, and 0.4 and very slowly for $ \omega_\mathrm{pmp} $=0.5. For $ \omega_\mathrm{pmp} $=0.3, 0.35, and 0.4, the rates are close to each other. For $ \omega_\mathrm{pmp} $=0.45, the rate at $ \omega_\mathrm{prb} < 0.06 $ is between those for $ \omega_\mathrm{pmp} $=0.3, 0.35, and 0.4 and that for $ \omega_\mathrm{pmp} $=0.5 [Fig.~\ref{fig:int_sgm}(a)], whereas the rate at $ \omega_\mathrm{prb} > 0.07 $ is close to those for $ \omega_\mathrm{pmp} $=0.3, 0.35, and 0.4 [Fig.~\ref{fig:int_sgm}(b)]. Namely, the number of carriers involved in the low-energy optical excitations is increased efficiently by $ 0.3 < \omega_\mathrm{pmp} < 0.4 $, but it is negligibly increased for $ \omega_\mathrm{pmp} $=0.5. 

This result shows that carriers introduced by photoexcitations with $ \omega_\mathrm{pmp} $ near 0.3 have low excitation energies and are regarded as delocalized. These excitations are characterized as interdimer CT excitations. Although any CT excitation weakens $ U_\mathrm{dim} $, it requires lattice motion and much time. Consequently, if a Mott-insulator-to-metal transition is induced, it is mainly through the introduction of carriers. On the other hand, carriers introduced by a photoexcitation with $ \omega_\mathrm{pmp} $=0.45 have finite excitation energies below the charge gap. A photoexcitation with $ \omega_\mathrm{pmp} $=0.5 introduces a negligible number of carriers. As a consequence, if a Mott-insulator-to-metal transition is induced, it is mainly through the weakening of $ U_\mathrm{dim} $. This excitation is characterized as an intradimer CT excitation. The characteristic of such a CT excitation seems to vary continuously as a function of the photoexcitation energy.

\section{Conclusions}

The present theoretical study is motivated by the photoinduced Mott-insulator-to-metal transition observed in the organic dimer Mott insulator, $ \kappa $-(BEDT-TTF)$_2$Cu[N(CN)$_2$]Br. \cite{kawakami_prl09} Using a two-dimensional three-quarter-filled extended Peierls-Hubbard model, we consider excitation-energy-dependent transition pathways. Bearing the above material in mind, we introduce quantum phonons that modulate intradimer transfer integrals so that the intradimer and the interdimer charge-transfer excitations have broad bands that overlap to a large extent. The transient quantities are obtained from the numerical solution to the time-dependent Schr\"odinger equation for the exact many-electron-phonon wave function on a small cluster. They indeed depend strongly on the excitation energy. 

The spectral-weight analysis shows that the number of delocalized carriers is increased efficiently when the excitation energy is around the peak in the conductivity spectrum.  If the system were large enough, this would lead to a photoinduced Mott-insulator-to-metal transition mainly through band-filling control. On the other hand, when the excitation energy is away from the peak on the high-energy side, few delocalized carriers are introduced. Slow lattice motion is induced by any CT excitation and modifies intradimer transfer integrals and weakens the effective interaction. Consequently, if a Mott-insulator-to-metal transition is induced by such a photoexcitation, it is mainly through bandwidth (relative to the effective interaction strength) control. Namely, photoexcitation-energy-dependent pathways are realized from the fact that, although the effective interaction is always and slowly weakened, the introduction of carriers is sensitive to the photoexcitation energy and proceeds much faster.

\section*{Acknowledgment}

This work was supported by Grants-in-Aid for Scientific Research (C) (Grant No. 19540381 and No. 23540426), Scientific Research (B) (Grant No. 20340101) and Scientific Research (A) (Grant No. 23244062), and by ``Grand Challenges in Next-Generation Integrated Nanoscience" from the Ministry of Education, Culture, Sports, Science and Technology of Japan, and the NINS program for cross-disciplinary study (NIFS10KEIN0160).

\bibliography{kappa11b_2}

\end{document}